\documentclass{svjour3}                    
\smartqed 
 \usepackage{graphicx,epsfig, cite}
\usepackage{fix-cm}
\usepackage{amsmath}

\def\be{\begin{equation}}
\def\ee{\end{equation}}

\def\ba{\begin{eqnarray}}
\def\ea{\end{eqnarray}}

\journalname{International Journal of Theoretical Physics}

\begin{document}

\title{ Lie symmetries for a conformally flat radiating star }

\author{G. Z. Abebe     \and K. S Govinder \and S. D. Maharaj}

\institute{
G.Z. Abebe \and K.S. Govinder        \and S.D. Maharaj \at
Astrophysics and Cosmology Research Unit,
 School of Mathematics, Statistics and Computer Science, Private Bag X54001,
 University of KwaZulu-Natal, Durban 4000, South Africa
\\
\email{maharaj@ukzn.ac.za}  \\ }
\date{Received: date / Accepted: date}

\maketitle

\begin{abstract}
We consider a relativistic  radiating spherical star in conformally flat spacetimes. In particular we study the junction condition relating the  radial pressure to the heat flux at the boundary of the star which  is a nonlinear  partial differential equation.   The Lie symmetry generators that leave the equation invariant are identified and we generate an optimal system. Each element of the optimal system is used to reduce the partial differential equation to an ordinary differential equation which is further analysed. We identify new categories of exact solutions to the boundary conditions. Two classes of solutions are of interest. The first class depends on a self similar variable. The second class is separable in the spacetime variables.


\end{abstract}

\section{Introduction}
Spherically symmetric spacetimes with nonvanishing heat flux are used in the modelling of radiating stars and in describing  cosmological models. The dynamics of gravitational collapse  with heat flow, as a highly dissipative process, has been extensively  studied by Herrera et al.~\cite{1,2}  and Di Prisco et al.~\cite{3} in both the free-streaming and diffusion approximations in relativistic astrophysics. Radiating models in cosmology are important in applications involving evolution of voids, formation of large scale structures and the study of singularities as discussed by  Krasinski \cite{4}. It is clear that heat flow is a vital ingredient in the modelling process. For  a relativistic radiating star the interior spacetime of the star  should match with the exterior Vaidya \cite{12} solution;  the junction condition relating the radial pressure with the heat flux at the boundary of the star must be satisfied. The full junction conditions for a radiating star were first generated by Santos \cite{16}. Such models are  important in investigating the physical features of radiating stars including dynamic stability, surface luminosity, relaxation effects, temperature  profiles and particle production at the boundary.

The assumption of conformal flatness is often made to integrate the Einstein field equations. Particular models discussing physical features of radiating spacetimes have been generated by Som and Santos \cite{10}, Maiti \cite{5}, Sanyal and Ray \cite{6}, Modak \cite{7}, Deng \cite{8} and Deng and Mannheim \cite{9}. The treatment of Ivanov \cite{11} is a global analysis of shear-free perfect fluid spheres with heat flow  containing as a special case  the condition of conformal flatness. Analytical models of radiating spherical gravitational collapse were studied by Grammenos and Kolassis \cite{13a} assuming conformal flatness and anisotropy in the pressure due to neutrino flow. With a vanishing Weyl tensor, Herrera et al.~\cite{13}  proposed a conformally flat relativistic model without solving the junction condition exactly. For this model Maharaj and Govender \cite{14} and  Herrera et al.~\cite{1}    subsequently generated exact classes of solutions by solving the junction condition directly in terms of elementary functions. Misthry et al.~\cite{15} generated other classes of solution with vanishing shear by transforming the junction condition equation to  an Abel equation. These new conformally flat solutions are useful in determining the gravitational behaviour of stars.

 The main objective of this paper is to  generate exact solutions for the equation governing the boundary condition of a conformally flat  radiating star. The Lie theory of differential equations is used to study the boundary conditions and to  generate exact solutions to the field equations.  This approach has been utilized very successfully in determining solutions to the Einstein field equations  \cite{a17,b17, c17, d17,e17,f17,g17,h17}.  Equations that have previously been intractable or solved in an {\it ad hoc} manner are able to be systematically analysed via this group theoretic method. We believe that the analysis here is the first comprehensive treatment of the boundary condition (with conformal flatness) using a symmetry approach. In this study, we do not impose the condition of pressure isotropy so that anisotropy is present. Note that the condition of pressure isotropy applies in most  earlier treatments which places further restrictions on the gravitational potentials since an additional differential equation has to be satisfied. Anisotropic pressures, where radial and tangential components arise, allow for more general behaviour.

  In \S \ref{335} we briefly discuss the conformally flat spacetime and  present the junction condition for a   radiating star. This equation is a highly nonlinear partial differential equation and difficult to solve  directly using traditional methods. Therefore we utilise a geometric approach to find solutions. In \S \ref{sec22} we obtain the Lie point symmetries  for the junction boundary  condition. Particular  symmetries  or any linear combination may   help us to obtain group invariant solutions. We transform  the boundary condition to an ordinary differential equation for each symmetry in the optimal system and exact solutions are found. In \S \ref{444} we find solutions invariant under elements of the optimal system. New models are generated as a result. Another new model,  this time invariant under the combination of symmetries $bG_1+G_3$, is given in \S \ref{666}. An analysis of the physical features indicates the model is reasonable close to the centre. In \S \ref{777} we make concluding remarks. 
  
  \section{The model\label{335}}
 We consider the particular case of spherically symmetric, shear-free spacetimes which are conformally flat when modelling  the interior of  a relativistic star. In this case there exists coordinates  $(x^a)=(t,r,\theta,\phi)$  for which the line element may be expressed in the form 
 \be\label{9}
 ds^2=B^2\left(-dt^2+  dr^2+r^2d\Omega^2 \right)  
 \ee
 where the metric function $B$ is a function of  $t$ and  $r$ and $d\Omega^2= d\theta^2+\sin ^2\theta d\phi^2 $. 
 
 The energy momentum tensor is given as
 \begin{equation}\label{7aa}
 T_{ab}=\left(\mu+p \right) u_au_b+pg_{ab}+q_au_b+q_bu_a+\pi _{ab}
 \end{equation}
where $\mu$ is the density, $p$ is the isotropic pressure, $q_a$ is the heat flux and $\pi_{ab}$ is the anisotropic stress. The stress tensor is 
\be 
\pi_{ab}=\left(  p_{\parallel}-p_{\perp}\right)\left(n_an_b-\frac{1}{3}h_{ab} \right) 
\ee
where $ p_{\parallel}$ is the radial pressure, $p_{\perp}$  is the tangential pressure and \textbf{n} is a unit radial vector. The isotropic pressure $p=\frac{1}{3}\left( p_{\parallel}+2p_{\perp} \right)$ relates the radial and the tangential pressures. The fluid four-velocity \textbf{u} is comoving and is given by
 $$u^a=\frac{1}{B}\delta^a_0$$  The heat flow  vector \textbf{q} takes the form  $$q^a=(0,q,0,0)$$  since $q^au_a=0$ and the heat is assumed to flow in the radial direction.
 The kinematical quantities for the line element \eqref{9} are given by
 \begin{subequations}
 \begin{eqnarray}
 \dot{u}^a&=&\left( 0,\frac{B_r}{B^3},0,0\right) \\
 \Theta &=&3\frac{B_t}{B^2}
 \end{eqnarray}
 \end{subequations}
 where $\dot{u}^a$ is the four-acceleration vector and $\Theta$ is the expansion scalar.
 
The Einstein field equations for the interior matter distribution become
\begin{subequations}\label{14}
 \begin{eqnarray}
 \mu&=&3\frac{B^2_t}{B^4}-\frac{1}{B^2}\left(2\frac{B_{rr}}{B}-\frac{B_r^2}{B^2}+\frac{4B_r}{rB} \right) \\
 p_{\parallel}&=&\frac{1}{B^2}\left( -2\frac{B_{tt}}{B}+\frac{B_t^2}{B^2}+ 3\frac{B^2_r}{B^2}+\frac{4}{r}\frac{B_r}{B}\right)\label{14p} \\
 p_{\perp}&=&-2\frac{B_{tt}}{B^3}+\frac{B_{t}^2}{B^4}+\frac{2}{r}\frac{B_r}{B^3}-\frac{B_r^2}{B^4}+2\frac{B_{rr}}{B^3} \label{14r}\\
 q&=&-\frac{2}{B^3}\left(-\frac{B_{rt}}{B} +2\frac{B_rB_t}{B^2}\right)\label{14d} 
 \end{eqnarray}
 \end{subequations}
for the line element \eqref{9}.
Equations \eqref{14} describe the  gravitational interactions in the  interior of a conformally flat star with heat flux and anisotropic pressure.

The boundary of a relativistic  radiating star divides the entire spacetime into two distinct regions: the interior spacetime and the exterior spacetime. The exterior radiating spacetime 
\be\label{va1}
ds^2=-\left(1-\frac{2m(v)}{R} \right)dv^2-2dvdR+R^2d\Omega^2
\ee
where $m(v)$ denotes the mass of the star as measured by an observer at infinity, was first derived by Vaidya \cite{12}. The spacetime is the unique spherically symmetric solution of the Einstein field equations for radially directed coherent radiation in the form of a null fluid. 
 The interior spacetime \eqref{9} has to be matched along the boundary of the star to this exterior Vaidya spacetime.  

The matching of the line elements \eqref{9} and  \eqref{va1}, and the matching of the  extrinsic curvature are necessary at the surface of the star. This matching leads to the following junction conditions
\begin{subequations}\label{rev1}
\begin{eqnarray}
Bdt&=&\left[ \left(1-\frac{2m}{R_\Sigma}+2\frac{dR_\Sigma}{dv} \right)^\frac{1}{2}dv\right] _\Sigma \\
\left(rB \right)_{\Sigma}&=&R_{\Sigma }\\
m(v)&=&\left[\frac{r^3}{2}\left( \frac{B_t^2}{B}-\frac{B_r^2}{B}\right)  -r^2B_r\right] _{\Sigma} \\
( p_{\parallel})_{\Sigma}&=&(Bq)_{\Sigma}
\end{eqnarray}
\end{subequations}
where  $\Sigma $ is the hypersurface that defines the boundary of the radiating sphere. The junction conditions \eqref{rev1} were completed by Santos \cite{16}. The particular junction condition 
\begin{equation}\label{15}
( p_{\parallel})_{\Sigma}=(Bq)_{\Sigma}
\end{equation}
is an additional differential equation  that has to be solved together with the interior field equations \eqref{14} to complete the model of a relativistic radiating star. This is a nonlinear differential equation which has to be integrated on the boundary $\Sigma$ of the star. By substituting equations \eqref{14p} and \eqref{14d} into \eqref{15} we have
\begin{eqnarray}\label{17}
2rBB_{rt}+2rBB_{tt}-4rB_rB_t-rB_t^2-3rB_r^2-4BB_r=0
\end{eqnarray}
at the boundary of a conformally  flat star. Equation \eqref{17} is the  master equation that governs the evolution of the model. We will attempt to integrate equation \eqref{17} using the Lie theory of extended groups applied to differential equations.

\section{Lie symmetry analysis\label{sec22}}
We use the Lie analysis in an attempt to find new solutions to \eqref{17}. We note that, except for Govinder and Govender \cite{16a}, no other attempt to apply symmetry analysis to the junction condition has been attempted. We know that an $n$th order differential equation 
\be 
E\left(r,t,B,B_r,B_{t},B_{rr},B_{rt},B_{tt}, \dots\right) =0
\ee
where $B=B(r,t)$,  admits a Lie point symmetry 
\be \label{sym10}
G=\xi_1\left(r,t,B \right)\frac{\partial}{\partial r}+\xi_2\left(r,t,B \right)\frac{\partial}{\partial t} +\eta\left(r,t,B \right)\frac{\partial}{\partial B}
\ee   
provided that
\be
\left. G^{[n]}E\right|_{E=0}=0 
\ee
where $G^{[n]}$ is the $n$th extension of the symmetry $G$ in \eqref{sym10} \cite{16c,16b}.
The method is algorithmic and
can be computed by using various software packages. Utilising \texttt{PROGRAM LIE} \cite{16h}, we can demonstrate that \eqref{17} admits the following
Lie point symmetries:
\begin{subequations}\label{18}
\begin{eqnarray}
G_1&=&\frac{\partial}{\partial t}\label{1cc}\\
G_2&=&t\frac{\partial}{\partial t}+r\frac{\partial}{\partial r}\label{3cc}\\
G_3&=&B\frac{\partial}{\partial B}\label{2cc}
\end{eqnarray}
\end{subequations}
with the nonzero Lie bracket relationship
$\left[ G_1,G_2\right]=G_1. $  
 \subsection{Optimal system}
 Given that  \eqref{17} has the three symmetries \eqref{18},  note that we can generate group invariant solutions using each symmetry in turn, or taking any linear combination of symmetries. Taking all possible combinations into account is not helpful. We proceed in a systematic manner by considering a subspace of this vector space. We utilize the subalgebraic structure of the symmetries \eqref{18} of the equation \eqref{17} to generate an optimal system of one-dimensional subgroups. Such an optimal system of subgroups is constructed by classifying the orbits of the infinitesimal adjoint representation of the Lie group on its related Lie algebra; this is achieved by using its infinitesimal generators. All group invariant solutions can be transformed to those obtained via this optimal
system \cite{16b}. 

The process  is algorithmic. To determine the inequivalent  subalgebras we begin with the following nonzero vector
\begin{equation}\label{20}
G = a_1 G_1 + a_2 G_2 + a_3 G_3
\end{equation}
 We try to remove as many of the coefficients, $a_i$ of $G$, as possible through judicious applications of adjoint maps to $G$. 
As a result, we have 
\begin{subequations}\label{mar7}
\begin{eqnarray}
G_1&=&\frac{\partial}{\partial t}\label{11cc}\\
G_2&=&t\frac{\partial}{\partial t}+r\frac{\partial}{\partial r}\label{13cc}\\
aG_2+G_3&=&a\left( t\frac{\partial}{\partial t}+r\frac{\partial}{\partial r}\right) +B\frac{\partial}{\partial B}\label{12cc}
\end{eqnarray}
\end{subequations}
are the subalgebra of the symmetries in \eqref{18}. 
\section{Solutions via symmetries in the optimal system\label{444}}
Using the generator
\be 
G_1=\frac{\partial}{\partial t}
\ee
 we determine the invariants from the invariant surface
condition
\begin{equation}
\frac{dt}{1}=\frac{dr}{0}=\frac{dB}{0}
\end{equation}
We obtain the invariants $r$ and 
\begin{equation}
B=y(r)
\end{equation}
for the generator $G_1$.
With  this transformation equation \eqref{17} is reduced  to 
\begin{equation}\label{redu11}
3ry'+4y=0
\end{equation}
Equation \eqref{redu11} is a first order, separable ordinary differential equation with solution
\begin{equation}\label{cccc}
y=\frac{1}{r^{4/3}} \quad \Rightarrow \quad B=\frac{1}{r^{4/3}}
\end{equation}
where   the constant of integration is taken to be unity.  Since the  gravitational potential $B$ in \eqref{cccc} is independent of time, this solution  cannot be applied to a radiating star.

The symmetry $aG_2+G_3$ transforms the master equation to 
 \begin{eqnarray}\label{mar2}
&&2a^2\left(x-1 \right)yy''+2a\left(1+a-3x-2ax \right)yy'+a^2\left(1-4x+3x^2 \right) y'^2\nonumber\\
&&+\left( 3+4a\right)y^2    =0
\end{eqnarray}
 which is a second order nonlinear ordinary differential equation. This equation is quite   difficult to solve since it has no symmetry in general for further reduction. 

The invariants of $G_2$ are given by 
\begin{subequations}\label{inv1}
\begin{eqnarray}
x&=&\frac{t}{r}\\
B&=&y\left(x\right)
\end{eqnarray}
\end{subequations}
For  this transformation equation \eqref{17} is  reduced to
\begin{equation}\label{3int}
\left( 2x-2\right)y y''+\left( 2-4x\right) yy'+\left(1-4x+3x^2 \right)y'^2 =0
\end{equation}
Equation \eqref{3int} is a second order nonlinear ordinary differential equation.  The integration of \eqref{3int} is not easy to complete  using traditional methods. However, using the computer package \texttt{MATHEMATICA} \cite{18} we find  the solution
\begin{equation}\label{1int}
y=c_2\exp\left( {\int_1^x\frac{8e^{2z}\left(z-1 \right) }{c_1+3e^{2z}-10ze^{2z}+6z^2e^{2z}}dz}\right) 
\end{equation}
We can simplify \eqref{1int} for particular parameter values by setting $c_1=0$ and $c_2=1$.
Noting that \eqref{17} is invariant under scalings of $B$, we obtain the particular solution
\be \label{1aa}
B= \left(\sqrt{7}+5-6\frac{t}{r}\right)^{\frac{14-2\sqrt{7}}{21}} \left(\sqrt{7}-5+6\frac{ t}{r}\right)^{ \frac{14+2\sqrt{7}}{21}} 
\ee
for the master equation \eqref{17}. 

We emphasize that the result \eqref{1aa} is a new exact solution to the boundary condition \eqref{15} for a conformally  radiating star with a shear-free matter distribution. It is not contained in any of the classes of solution found in previous investigations. The elementary form of the solution in \eqref{1aa}  will assist in studying the physical features of a conformally flat radiating star. 

 For the solution \eqref{1aa} the line element \eqref{9} becomes
 \begin{equation}\label{4.2b}
 ds^2=\left[\left(\sqrt{7}+5-6\frac{ t}{r}\right)^{\frac{28-4\sqrt{7}}{21}} \left(\sqrt{7}-5+6\frac{t}{r}\right)^{\frac{28+4\sqrt{7}}{21}}\right] \left(-dt^2+  dr^2+r^2d\Omega^2 \right)  
 \end{equation}
  The kinematical quantities for the line element \eqref{4.2b} are given by
\begin{subequations}
\begin{eqnarray}
\dot{u}^a&=&\left( 0,\frac{2 r^3 (r-t) t \left(5+\sqrt{7}-6\frac{ t}{r}\right)^{\frac{4}{3 \sqrt{7}}}  \left(\frac{12 (5 r-3 t) t}{r^2}-18\right)^{2/3}}{9 \left(3 r^2-10 r t+6 t^2\right)^3\left(\sqrt{7}-5+6\frac{ t}{r}\right)^{\frac{4}{3 \sqrt{7}}}},0,0\right) \\
&&\nonumber\\ 
\Theta&=&\frac{4\sqrt[3]{6}  (r-t) \left(5+\sqrt{7}-6\frac{ t}{r}\right)^{\frac{2}{3 \sqrt{7}}} }{\left(\sqrt{7}-5+6\frac{ t}{r}\right)^{\frac{2}{3 \sqrt{7}}}\left(3-2 (5 r-3 t)\frac{t}{r^2}\right)^{2/3} \left(3 r^2-10 r t+6 t^2\right)}
\end{eqnarray}
\end{subequations}
We note that both the acceleration and expansion grow smaller with increasing time. The spacetime approaches asymptotic flatness.

 The matter variables are given by
\begin{subequations}\label{field1}
\begin{eqnarray}
 \mu&=&\frac{8 \left(5+\sqrt{7}-6\frac{ t}{r}\right)^{\frac{4}{3 \sqrt{7}}} \left(-5+\sqrt{7}
+6\frac{ t}{r}\right)^{-\frac{4}{3 \sqrt{7}}} \left(12 r^4-24 r^3 t+15 r^2 t^2-4 r t^3+
2 t^4\right)}{3 \left(18-\frac{12 (5 r-3 t) t}{r^2}\right)^{1/3}
 \left(3 r^2-10 r t+6 t^2\right)^3}\nonumber\\
   &&\\
  p_{\parallel}&=&\frac{8 r \left(5+\sqrt{7}-6\frac{ t}{r}\right)^{\frac{4}{3 \sqrt{7}}} \left(-5+\sqrt{7}+6\frac{ t}{r}\right)^{-\frac{4}{3 \sqrt{7}}} \left(3 r^3+2 r^2 t-12 r t^2+8 t^3\right)}{3 \left(18-\frac{12 (5 r-3 t) t}{r^2}\right)^{1/3} \left(3 r^2-10 r t+6 t^2\right)^3}\nonumber\\
 &&\\
 p_{\perp}&=&\frac{4\sqrt[3]{4} (r-t) \left(5+\sqrt{7}-6\frac{ t}{r}\right)^{\frac{4}{3 \sqrt{7}}}
 \left(-5+\sqrt{7}+6\frac{ t}{r}\right)^{-\frac{4}{3 \sqrt{7}}} \left(3 r^3-4 r^2 t+8 r t^2-4 t^3\right)}{3 
\left(9-\frac{6 (5 r-3 t) t}{r^2}\right)^{1/3} \left(3 r^2-10 r t+6 t^2\right)^3}
\nonumber\\
&&\\
q&=&\frac{4 r^3 \left(5+\sqrt{7}-6\frac{ t}{r}\right)^{\frac{2}{\sqrt{7}}} \left(-5+\sqrt{7}+6\frac{ t}{r}\right)^{-\frac{2}{\sqrt{7}}}
 \left(3 r^3+2 r^2 t-12 r t^2+8 t^3\right)}{9 \left(3 r^2-10 r t+6 t^2\right)^4}\nonumber\\
 &&
\end{eqnarray}
\end{subequations}
for the metric \eqref{4.2b}.

We remark that this new solution is given in terms of a self-similar variable $x=t/r$. The appearance of the self-similar variable implies the existence of a homothetic Killing vector. In shearing spherically symmetric spacetimes a homothetic vector was found by Wagh and Govinder \cite{17}. The full conformal geometry of both shear-free and shearing spacetimes was completed by Moopanar and Maharaj \cite{mama2, mama1}, respectively.

\section{Invariance under $bG_1+G_3$ \label{666}} 
 Here we consider the combination $bG_1+G_3$ which is not in the optimal system. This approach is taken as we were not able to solve all the equations obtained via the optimal system.  It turns out that this is the best combination that yields a solution.  In the symmetry
  \begin{equation}\label{gez1}
bG_1+G_3=b\frac{\partial}{\partial t}+B\frac{\partial}{\partial B}
\end{equation} 
the constant  $b$ is  nonzero  and arbitrary. 
For the symmetry \eqref{gez1}, we determine the  invariants $r$ and
\begin{equation}\label{gez2}
B=\exp\left(\frac{t}{b}\right) y(r)
\end{equation}

Using this transformation, equation \eqref{17} is reduced to the first order ordinary  differential equation
\be\label{gez3}
3 b^2 r y'^2+2 b (2 b+r) y y'-r y^2=0
\ee
This is a highly nonlinear equation and is  difficult to solve. However, equation \eqref{gez3} can be integrated with the help of  \texttt{MATHEMATICA}  \cite{18} to give  two special solutions
\begin{subequations}\label{gez5}
\begin{eqnarray}
y&=& \frac{\sqrt[3]{b+2r+2f(r)  }}{\exp\left(\frac{r-2f(r)}{3b} \right)\sqrt[3]{\left[ 2b+r+2f(r)\right]^2 }} \\  
y&=&\frac{\sqrt[3]{\left( 2b+r+2f(r)\right)^2 }}{r^{4/3}\exp\left(\frac{r+2f(r)}{3b} \right)\left(\sqrt[3]{ b+2r+2f(r)   } \right) }
\end{eqnarray}
\end{subequations}
where the  constants of integration are set to unity and $f(r)= \sqrt{b^2+br+r^2}$. Hence we have found particular solutions to \eqref{17} of the form
\begin{subequations}\label{gez7} 
\begin{eqnarray}
B&=& \exp\left(\frac{3t-r+2f(r)}{3b} \right)\left( \frac{ b+2r+2f(r)  }{\left[ 2b+r+2f(r)\right]^2 }\right) ^{1/3}\label{qqq2} \\
B&=&   \exp\left(\frac{3t-r-2f(r)}{3b} \right)\left( \frac{\left[ 2b+r+2f(r)\right]^2 }{   r^{4}\left[ b+2r+2f(r)\right]   }\right) ^{1/3}  \label{qqq1}   
\end{eqnarray}
\end{subequations}
Note that in the above  solutions $b\ne 0$ or we would simply have $B=B(r,t)$ in \eqref{gez2}. 

We have found two new solutions to the boundary condition \eqref{15} for  a radiating star. The solutions \eqref{qqq2} and \eqref{qqq1} have been generated using invariance under $bG_1+G_3$ which is not in the optimal system. The metric function $B$ is separable in the variables $t$ and $r$ in this class of solutions. We point out the fact that the new solutions are given in terms of elementary functions  and this will help in the analysis of the physical features of a stellar model. The solution \eqref{qqq2} has the desirable feature of being regular at the stellar centre but the heat flux has the form
\be 
q=-\frac{2r e^{\frac{r-2 f(r)-3 t}{b}}  \left(2 b+r+2 f(r)\right)}{b^2 \left(b+2 \left(r+f(r)\right)\right)}
\ee
The heat flux is always negative for positive $b$ which implies inflow of energy across the boundary of the star. For a realistic model of radiating body the heat flow should be outwards to the exterior. This example suggests that even though Lie analysis does provide new solutions to the boundary condition, a careful analysis of the physical features is still necessary. The solution \eqref{qqq1} has a singularity at $r=0$ and can only be applied in regions away from the stellar centre. Close to the centre another solution has to used; solution \eqref{qqq1} should be used as part of core-envelope model for radiating star. Solution \eqref{qqq1} has several desirable features which become clear in our physical analysis for regions away from the  singularity at the centre of the radiating star.   This realistic solution  may be helpful in describing the  interior spacetime of a  radiating star in conformally flat spacetimes.

 For the solution \eqref{qqq1} the line element \eqref{9} becomes
 \begin{eqnarray}\label{4.20bbb}
 ds^2&=&\left[\exp\left(\frac{3t-r-2f(r)}{3b} \right)\left( \frac{\left[ 2b+r+2f(r)\right]^2 }{   r^{4}\left[ b+2r+2f(r)\right]   }\right) ^{1/3} \right] ^2\nonumber\\
 && \times \left(-dt^2+  dr^2+r^2d\Omega^2 \right)  
 \end{eqnarray}
  where $0\le t\le \infty$. The kinematical quantities are  given by
 \begin{subequations}\label{zzz1}
\begin{eqnarray}
\dot{u}^a&=&\left(0,\frac{e^{\frac{2 \left(r+2 f(r)-3 t\right)}{3 b}} r^{5/3}}{-b f(r) \left(2 b+r+2 f(r)\right)^{7/3} \left(b+2 \left(r+f(r)\right)\right)^{1/3}}\left(8 b^4+6 r^3 \left(r+f(r)\right)\right.\right.\nonumber\\
&& \left. \left.+4 b^3 \left(5 r+2 f(r)\right)+3 b r^2 \left(6 r+5 f(r)\right)+2 b^2 r \left(13 r+8 f(r)\right)\right),0,0\right) \\
&&\nonumber\\
\Theta&=&\frac{3 e^{\frac{r+2 f(r)-3 t}{3 b}} r^{4/3} \left(b+2 \left(r+f(r)\right)\right)^{1/3}}{b \left(2 b+r+2 f(r)\right)^{2/3}}
\end{eqnarray}
\end{subequations}
for the line element \eqref{4.20bbb}. The  acceleration and the expansion decrease for increasing time and the spacetime becomes asymptotically flat.

\begin{figure}[h!]
\centering
\includegraphics[scale=.5]{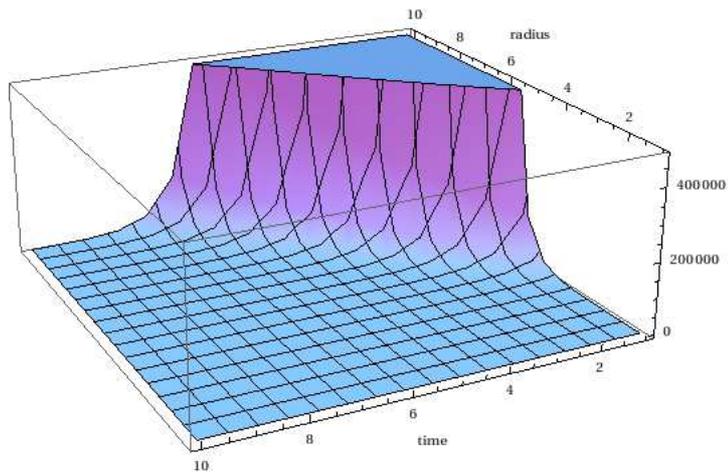} \\
\caption{Energy density \label{new1}}
\end{figure}

\begin{figure}[h!]
\centering
\includegraphics[scale=.5]{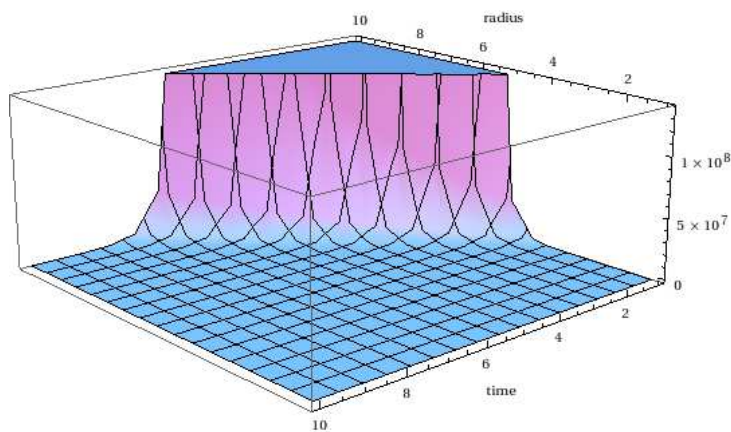} \\
\caption{Heat flux \label{new2}}
\end{figure}

  The matter variables become
  \begin{subequations}
 \begin{eqnarray}
 \mu&=& \frac{2e^{\frac{2 \left(r+2 f(r)-3 t\right)}{3 b}} r^{2/3}}{b^2 f(r) \left(2 b+r+2 f(r)\right)^{10/3} \left(b+2 \left(r+f(r)\right)\right)^{4/3}}\left(32 b^7+72 r^6 \left(r+f(r)\right) \right.\nonumber\\
&&\left.+16 b^6 \left(9 r+2 f(r)\right)+36 b r^5 \left(9 r+8 f(r)\right)+25 b^4 r^2 \left(31 r+14 f(r)\right)+18 b^3 r^3 \right.\nonumber\\
&&\left.\times \left(51 r+31 f(r)\right)+9 b^2 r^4 \left(78 r+59 f(r)\right)+2 b^5 r \left(213 r+64 f(r)\right)\right)\\
  p_{\parallel}&=&\frac{6 e^{\frac{2 \left(r+2 f(r)-3 t\right)}{3 b}} r^{5/3}}{b^2 \left(2 b+r+2 f(r)\right)^{10/3} \left(b+2 \left(r+f(r)\right)\right)^{4/3}}\left(32 b^5+24 r^4 \left(r+f(r)\right)\right.\nonumber\\
&&\left.+16 b^4 \left(7 r+2 f(r)\right)+12 b r^3 \left(8 r+7 f(r)\right)+6 b^3 r \left(31 r+16 f(r)\right)\right.\nonumber\\
&&\left.+3 b^2 r^2 \left(59 r+42 f(r)\right)\right)\\
 p_{\perp}&=&\frac{6 e^{\frac{2 \left(r+2 f(r)-3 t\right)}{3 b}} r^{2/3}}{f(r) \left(2 b+r+2 f(r)\right)^{16/3} \left(b+2 \left(r+f(r)\right)\right)^{7/3}} \left(1024 b^8+648 r^7 \left(r+f(r)\right)\right.\nonumber\\
&&\left.+5376 b^6 r \left(3 r+f(r)\right)+256 b^7 \left(23 r+4 f(r)\right)+324 b r^6 \left(13 r+12 f(r)\right)\right.\nonumber\\
&&\left.+81 b^2 r^5\left(160 r+133 f(r)\right)+81 b^3 r^4 \left(303 r+220 f(r)\right)+16 b^5 r^2 \right.\nonumber\\
&&\left.\times\left(1709 r+816 f(r)\right)+4 b^4 r^3 \left(7807 r+4748 f(r)\right)\right)\\
q&=&\frac{2}{9 b^2} e^{\frac{r+2 f(r)-3 t}{b}} r \left(2 b^2+2 r \left(r+f(r)\right)-b \left(r+2 f(r)\right)\right)
 \end{eqnarray}
 \end{subequations} 
 for the metric \eqref{4.20bbb}. The quantities $\mu$, $ p_{\parallel}$, $p_{\perp}$ and $q$ are regular in the interior of the relativistic star. They remain continuous and well behaved in regions of spacetime surrounding the stellar core. This feature is illustrated in Fig. \ref{new1}--\ref{new2}. Fig. \ref{new1} is a plot for energy density $\mu$ and Fig. \ref{new2} illustrates the heat flow $q$; both plots represent regular behaviour over the interval. We have omitted plots for $ p_{\parallel}$ and $ p_{\perp}$ as they represent profiles which are similar to that of the energy density $\mu$.  These are desirable features and point to a physically viable model. We note that the  heat flux decreases for increasing time. This implies that the star is radiating away energy as it approaches a static limit.
\section{Conclusion \label{777}}
We  considered a relativistic  radiating star in conformally flat spacetimes. We studied the junction condition  which relates the radial pressure to the heat flux  which is  the master equation. We demonstrated that this equation admits three Lie point symmetries and obtained an optimal system. This was  used to reduce the governing highly nonlinear partial differential equation  to  ordinary differential equations. We also used a symmetry combination which was not in the optimal system.  By solving the reduced ordinary differential equations and transforming to the original variables we obtained new exact solutions for the master equation. We believe that the solutions obtained in this paper are not contained in the literature. Two classes of models are of particular interest. The first class depends on the self similar variable $t/r$. In the second class the metric function is separable in the spacetime variables $t$ and $r$. Two particular metrics could be identified in the second class. The first metric is regular at the centre but the heat flow is inwardly directed. The second metric is not regular at the centre but the heat flow is outwardly directed. Clearly the Lie analysis of differential equations is a useful technique in generating exact solutions to the boundary condition. However a subsequent study of the physical features remains necessary. For our example the matter variables are regular in a spacetime region at least close to the centre. The heat flux, acceleration and expansion are decreasing functions for large time. In our example observe that we obtain the relationship
\be 
   p_{\parallel}=\mu\lambda,
   \quad \lambda = \frac{3 r f(r)}{5 b^2+8 b r+8 r^2-4 b f(r)-5 r f(r)}
   \ee
   which relates the $ p_{\parallel}$ and $\mu$. Thus the ratio  $ \frac{p_{\parallel} }{ \mu}$ is independent of time. This property essentially arises from the separability of the metric \eqref{qqq1}. 

\begin{acknowledgements}
GZA and KSG thank the National Research Foundation,  African Institute for Mathematical Sciences and University of KwaZulu--Natal for continuing support.
SDM acknowledges that this work is based upon research supported by the South African Research Chair Initiative of the Department of Science and Technology and the National Research Foundation.
\end{acknowledgements}


\begin{thebibliography}{}
  \bibitem{1}
   Herrera, L., Di Prisco, A.,  Ospino, J.: { \it Phys. Rev. D} {\bf 74}, 044001 (2006)
  
  \bibitem{2} 
  Herrera, L.,  Ospino, J., Di Prisco, A.,  Fuenmayor, E.,   Troconis, O.: {\it J. Math. Phys.} {\bf 43}, 064025 (2009)
  
 \bibitem{3}  
  Di Prisco, A., Herrera, L.,  Le Denmat, G.,  MacCallum, M.A.H.,  Santos, N.O.: 
 {\it Phys. Rev. D} {\bf 74}, 064017 (2007)
 
  \bibitem{4}
  Krasinski, A.: Inhomogeneous Cosmological Models, Cambridge University Press, Cambridge (1997)
  
 \bibitem{12}
 Vaidya, P.C.: \emph{Proc. Ind. Acad. Sci. A}  {\bf 33}, 264 (1951)
 
  \bibitem{16} 
   Santos, N.O.: {Mon. Not. R. Astron. Soc.} {\bf 216}, 403 (1985)
 
 \bibitem{10}
 Som, M.M., Santos, N.O.: {\it Phys. Lett. A}  {\bf 87}, 89 (1981) 
 
 \bibitem{5}
 Maiti, S.R.: {\it  Phys. Rev. D} {\bf 25},  2518 (1982)
 
\bibitem{6}
 Sanyal, A.K.,  Ray, D.: {\it J. Math. Phys.} {\bf 25},  1975 (1984)

  \bibitem{7}
  Modak, B.:  {\it Astrophys. Astr.} {\bf 5}, 317 (1984)
 
  \bibitem{8}
   Deng, Y.: {\it Gen. Relativ. Gravit.}  {\bf 21}, 503  (1989)

  \bibitem{9}
   Deng, Y.,  Mannheim, P.D.:  {\it  Phys. Rev. D}  {\bf 42}, 371 (1990)
 
  \bibitem{11} 
  Ivanov, B.V.: {\it Gen. Relativ. Gravit.} {\bf 42},  1835 (2012)
  
  \bibitem{13a}
  Grammenos, T., Kolassis,  C.: {\it Phys. Lett. A} {\bf 169}, 5 (1992)
  
  
   \bibitem{13} 
   Herrera, L., Le Denmat, G., Santos, N.O., Wang, A.: {\it Int. J. Mod, Phys. D} {\bf 13},  583 (2004)
  
  \bibitem{14}
   Maharaj, S.D., Govender, M.: {\it Int. J. Mod. Phys. D} {\bf 14}, 667 (2005)
    
  \bibitem{15}
   Misthry, S.S., Maharaj, S.D., Leach, P.G.L.: {\it Math. Meth. Appl. Sci.} {\bf 31}, 363 (2008)   
 
  \bibitem{a17}
  Govinder, K.S.,   Leach, P.G.L.,  Maharaj, S.D.: {\it Int. J. Theor. Phys.} { \bf 34}, 625 (1995) 
 
  \bibitem{b17} 
  Leach, P.G.L.,   Govinder, K.S.: {\it Qu\ae st Math.} {\bf 19}, 163 (1996) 

 \bibitem{c17}
  Hansraj, S.,  Maharaj, S.D., Msomi, A.M.,  Govinder, K.S.: {\it J. Phys. A: Math. Gen.} {\bf 38}, 4419 (2005)

 \bibitem{d17}
  Msomi,  A.M.,    Govinder, K.S.,    Maharaj, S.D.: {\it J. Phys. A: Math. Theor.} { \bf 43}, 285203 (2010)
  
  \bibitem{e17}
   Kweyama, M.C.,  Govinder, K.S.,  Maharaj, S.D.: {\it Class. Quantum Grav.} {\bf  28}, 105005 (2011)
  
  \bibitem{f17}
  Msomi,  A. M., Govinder,  K. S., Maharaj, S.D.: { \it Gen. Relativ. Gravit.} { \bf 43}, 1685 (2011)
   
  \bibitem{g17} 
   Msomi,  A.M., Govinder, K.S.,    Maharaj, S.D.: { \it Int. J. Theor. Phys.} { \bf 51}, 1290 (2012)
  
   \bibitem{h17}
   Govinder,  K.S.,    Hansraj, S.: {\it J. Phys A: Math. Theor.} {\bf 45}, 155210 (2012)
  
  \bibitem {16a} 
  Govinder, K.S.,  Govender, M.A.: {\it Gen. Relativ. Gravit.} {\bf 44}, 147 (2012)

 \bibitem{16c} 
 Bluman, G.W.,  Cheviakov, A.F.,   Anco, S.C.: Applications of symmetry methods to partial differential  equations, Springer-Verlag, New York (2010)
 
 \bibitem{16b}
 Olver, P.J.: Applications of Lie Groups to Differential Equations, Springer-Verlag, New York (1993)
 
 \bibitem{16h}  Head, A.K.: {\it Comp. Phys. Comm.} {\bf 71},  241 (1993)

 \bibitem{18}  
 Wolfram, S.: \texttt{MATHEMATICA},  Wolfram Research, Champaign (2008)

 \bibitem{17}
 Wagh, S., Govinder, K.S.: {\it Gen. Relativ. Gravit.} {\bf 38},  1253  (2006)
 
 \bibitem{mama2}
  Moopanar, S., Maharaj, S.D.: {\it J. Eng. Math.}, in press (2013)    
 
  \bibitem{mama1}
 Moopanar, S., Maharaj, S.D.: {\it Int. J. Theor. Phys.} {\bf 49},  1879  (2010)  
 \end{thebibliography}
 \end{document}